\documentclass[fleqn,10pt]{wlscirep}

\usepackage{subfig}
\usepackage{geometry}
\usepackage{graphicx}
\usepackage{mathtools}
\usepackage{booktabs}
\usepackage{multicol}
\usepackage{multirow}
\usepackage{threeparttable}
\usepackage{float}
\usepackage{epstopdf}
\usepackage{rotating}
\usepackage{colortbl}

\definecolor{LightCyan}{rgb}{0.88,1,1}

\def\etal{\mbox{\it et al.\ }}

\def\ln{\textrm{ln}}

\newcommand{\fig}[1]{Fig.~\ref{#1}}

\newcommand{\tab}[1]{Table~\ref{#1}}

\title{On the stochastic phase stability of Ti$_2$AlC-Cr$_2$AlC }

\author[1,*]{Thien C. Duong}
\author[1]{Anjana Talapatra}
\author[1]{Woongrak Son}
\author[1,2]{Miladin Radovic}
\author[1,2]{Raymundo Arr\'{o}yave}

\affil[*]{terryduong84@tamu.edu}

\affil[1]{Department of Materials Science and Engineering, Texas A$\&$M University, College Station, TX 77843, United States}
\affil[2]{Department of Mechanical Engineering, Texas A$\&$M University, College Station, TX 77843, United States}


\begin{abstract}
	
The quest towards expansion of the M$_{n+1}$AX$_{n}$ design space has been accelerated with the recent discovery of several solid solution and ordered phases involving at least two M$_{n+1}$AX$_{n}$ end members. Going beyond the nominal M$_{n+1}$AX$_{n}$ compounds enables not only fine tuning of existing properties but also entirely new functionality. This search, however, has been mostly done through painstaking experiments as knowledge of the phase stability of the relevant systems is rather scarce. In this work, we report the first attempt to evaluate the finite-temperature pseudo-binary phase diagram of the Ti$_2$AlC-Cr$_2$AlC via first-principles-guided Bayesian CALPHAD framework that accounts for uncertainties not only in \textit{ab initio} calculations and thermodynamic models but also in synthesis conditions in reported experiments. The phase stability analyses are shown to have good agreement with previous experiments. The work points towards a promising way of investigating phase stability in other MAX Phase systems providing the knowledge necessary to elucidate possible synthesis routes for M$_{n+1}$AX$_{n}$ systems with unprecedented properties.
	
\end{abstract}

\begin{document}
	
\flushbottom
\maketitle
	
\section*{Introduction}

M$_{n+1}$AX$_{n}$, wherein M is an early transition metal, A is an A-group element, X is carbon or nitrogen, belong to a special class of nanolaminate materials. They possess hexagonal $P63/mmc$ structure within which M-X layers interleave with A layers. This structure allows the coexistence of both metallic (M-A) and covalent/ionic (M-X) bonds which features a unique combination of ceramic- and metallic-like properties, e.g. high stiffness, good corrosion resistance, good conductivity, high damage tolerance, and machinability \cite{barsoum2000a, barsoum2004, 2011Barsoum, 2013Barsoum, 2013Radovic}. Thanks to this unique combination, M$_{n+1}$AX$_{n}$ are extremely promising for advanced high-temperature applications, and therefore of great interest.

The design of materials and/or components for advanced high-temperature applications based on MAX phases requires knowledge of physical and mechanical properties. Such knowledge could only be achieved via successful synthesis and characterization of the materials. So far, roughly over 70 MAX compounds and approximately 100 solid solutions have been synthesized. These are but a small fraction of the `pure' 600 MAX compounds \cite{2014Aryal} and billions (if not trillions) of solid solutions that could possibly exist, i.e. those that are elastic and thermodynamic stable. The discovered MAX phases, therefore, constitute only a small portion of the multi-dimensional MAX design space. Richer and better solutions for materials and/or components design based on MAX phases are to be expected within the remaining undiscovered MAX materials design space; and, to follow up, engineering questions such as ``what is the MAX phase with highest Young modulus?'' or ``among the MAX alloys which ones exhibit solid-solution strengthening?'' will naturally rise. In order to answer such questions, it is necessary: for (1) a systematic 
research and development scheme to be reasonably sketched according to a specific engineering problem and to integrate such scheme with (2) combined high-throughput synthesis and characterization capabilities \cite{2008ICME, 2016Lookman}. The latter, although experiencing a great cost inertia, has started to grow. This is especially true with the advent of computer-assisted schemes for materials development. Indeed, recent developments in computer infrastructure and simulation methods have enabled the accelerated development of high-throughput computational materials design.

Motivated by the increasing interest in MAX alloys and the engineering problem of solid-solution strengthening of MAX phases \cite{2016Gao, 2016Talapatra}, we have developed a design scheme based on high-throughput first-principles calculations. The scheme currently has two steps. The first step is to study the effect of M site alloys on the solid solutions of M$^1_{2}$AX-M$^2_{2}$AX system, where M$^1$ and M$^2$ are two different transition metals, using cluster expansion approach. Based on this work, suggestions on MAX phases that possibly exhibit solid solution with each other (and hence solution strengthening) will be made. The result of this work for the case of M$^1_{2}$AlC-M$^2_{2}$AlC has already been reported elsewhere \cite{2016Talapatra}. To follow the first step, the second step is to calculate the pseudo-binary phase diagram of the selected M$^1_{2}$AX-M$^2_{2}$AX system that may exhibit solid solution. The idea of a pseudo-binary diagram is interesting in the sense that when two pure MAX phases are brought together and their chemical reaction could be promoted such a diagram would specify the conditions under which solid solutions are likely to be formed---or whether such a state is not thermodynamically competitive with neighboring compounds in the materials design space. When the reaction is instead promoted via the use of pure elements or their end-member forms, which is usually the case, the pseudo-binary phase diagram could still provide a better understanding of phase formation. The current work is meant to elaborate on the second step within our design scheme, and the chosen system is Ti$_2$AlC-Cr$_2$AlC for its practical value as well as available experimental phase-equilibrium data.

Ti$_2$AlC and Cr$_2$AlC are promising candidates for oxidation resistant, autonomous self-healing materials \cite{2010Yu, 2012Li, 2013Tallman, 2011Basu, 2015Smialek}. This is due to the fact that the materials upon cracking will have Al, located near the crack area, react with oxygen and the resulting Al$_2$O$_3$ will fill in the crack space. The healing process via oxidation is observed to be very good in Ti$_2$AlC, with its fracture strength recovering almost to its original levels \cite{2012Li}. It is, however, found that Ti$_2$AlC, during oxidation, can also form TiO$_2$ which may serve as a crack-initiation site \cite{2012Li}. Compared to Ti$_2$AlC, although Cr$_2$AlC has a relatively slow healing rate, it does not tend to negatively impact the healing process upon oxidation and in addition exhibits a remarkable corrosion resistance at elevated temperatures \cite{2007Lin, 2008Tian, 2008Lee}. To achieve a good healing rate, avoid the formation of TiO$_2$, and increase corrosion resistance, (Ti,Cr)$_2$AlC alloys seem to be a reasonable solution. Ti$_2$AlC, Cr$_2$AlC, and their potential alloys are also excellent candidates for nuclear cladding materials. Indeed, the high oxidation and corrosion resistances of the MAX phases can greatly improve the inherent safety factor of nuclear reactors under operating and severe-incident conditions. The materials also exhibit low neutron absorption cross section \cite{2012Hoffman, 2015Horlait} which enhances the efficiency of nuclear reactors and hence their economic factor.

The study of Ti$_2$AlC -- Cr$_2$AlC phase equilibria has branches in both experimental and computational work. The first experimental study dates back to 1980 when Schuster \etal \cite{1980Schuster} investigated Cr$_2$AlC, Ti$_2$AlC, V$_2$AlC, and their possible solid solutions. They had tried to synthesize (Cr,Ti)$_2$AlC solution with various proportion of Cr$_2$AlC and Ti$_2$AlC, and identified that 6 at.\% maximum solubility of Cr in Ti$_2$AlC and about $25 at.\%$ solubility of Ti in Cr$_2$AlC. To follow Schuster \etal, Kim \etal \cite{2010Kim} and Lee \etal \cite{2010Lee} both tried to synthesize (Cr,Ti)$_2$AlC in 2010. In their work, the focus was on the Cr-rich side and they have concluded, based on X-ray diffraction analysis, that the maximum solubility of Ti in Cr$_2$AlC is somewhere between $10$ and $20$ $at.\%$, lower than that reported by Schuster \etal. Interestingly, Lee \etal \cite{2010Lee, 2012Lee}, while studying (Cr$_{0.95}$Ti$_{0.05}$)$_2$AlC, have pointed out that the solubility of Ti in Cr$_2$AlC is limited to only $5$ $at.\%$, raising an interesting fundamental question on the maximum solubility of Ti in Cr$_2$AlC. To follow, Ying \etal \cite{2014Ying} attempted to synthesis (Cr,Ti)$_2$AlC with compositions ranging from $12.5$ $at.\%$ to $75$ $at.\%$ but found only Cr$_2$AlC and Ti$_3$AlC$_2$-like phases, implying possibly a limit of $12.5$ $at.\%$ on the solubility of Ti in Cr$_2$AlC. In 2015, Horlait \etal \cite{2015Horlait} studied the Ti-Cr-Al-C system, focusing on the (Cr$_x$,Ti$_{1-x}$)$_2$AlC mixed compositions with $x$ = 0, 0.05, 0.25, 0.5, 0.75, 0.95 and 1. They have reported the observation of solid solution at Cr$_{0.05}$ and above this composition the coexistences of MAX phases, Al$_x$Cr$_y$ intermetallic compounds, and titanium carbides were found. Interestingly, among the observed MAX phases in the composite is the ordered (Cr$_{2/3}$,Ti$_{1/3}$)$_3$AlC$_2$, which was identified by Liu \etal \cite{2014aLiu, 2014bLiu} not very long before. Compared with the others, the work of Horlait et al., although still with limited information, is perhaps the most complete assessment of the Ti$_2$AlC-Cr$_2$AlC pseudo-binary phase diagram.

From the theoretical side, the first computation regarding the phase stability of (Cr,Ti)$_2$AlC was of Sun \etal \cite{2003Sun}. Inspired by the fact that solid solution is one efficient way to tune the properties of MAX phases, Sun \etal conducted a theoretical investigation of mutual substitution of Ti and Cr in M$_2$AlC, within the framework of density functional theory (DFT). Based on their electronic structure calculations, Sun \etal concluded that (Cr,Ti)$_2$AlC solid solution could be metastable. This is owed to the fact that the materials exhibit a small formation energy (almost flat) and that their $E_F$ lie either at the peak or fall in the pseudogap between bonding and antibonding states. Sun \etal, however, did not consider phase competition in their study, which is important for understanding phase equilibria of the system (as demonstrated in the aforementioned work of Horlait \etal \cite{2015Horlait}). Later, Keast \etal \cite{2009Keast} followed by Dahlqvist \etal \cite{2010aDahlqvist, 
2010bDahlqvist} had taken into account phase competition in their phase-stability studies, albeit their focus was more on the end-member systems rather than their mutual solution. One of the take-aways from both Keast and Dahlqvist \etal studies was the fact that MAX phases with higher $n$ values can get involved in the phase competition with $M_2AX$ phases and therefore the latter should be taken into account. The most recent computational work was of Shang \etal \cite{2014Shang} in which the authors studied the phase stability of (Cr$_{1-x}$,M$_x$)$_2$(Al$_{1-y}$,A$_y$)(C$_{1-z}$,X$_z$) with (M = Ti, Hf, Zr, A = Si, and X = B). Similar to Sun \etal, Liu \etal focus was on whether these phases are thermodynamically stable and hence overlooked the phase competition. From a general point of view, these theoretical studies have contributed to a better understanding of Ti$_2$AlC-Cr$_2$AlC phase equilibria from Gibb's energetic perspective (which guarantees the self-consistency between thermochemistry and phase equilibria). However, they are restricted to the ground-state condition, and hence interpretation of their insights regarding phase equilibria to advanced temperatures is somewhat limited.

In this work, we have attempted to evaluate the finite-temperature pseudo-binary phase diagram of Ti$_2$AlC-Cr$_2$AlC based on high throughput first-principles calculations and Bayesian CALPHAD. The effect of phase competition on equilibria at finite temperature was taken into account by considering the relative stability of possible unary, binary, ternary, and quaternary intermetallic compounds with respect to the MAX end members. The compounds were collected from previous literature, Ti-Cr, Ti-Al, Ti-C, Cr-Al, Cr-C, and Al-C phase diagrams. Their finite-temperature free energies were valuated, firstly using first-principles calculations which take into account both vibrational and electronic contributions to the total free energy. The energies were, in turn, `refined' within the framework of CALPHAD methodology, by introducing finite-temperature phase-equilibria constraints available from experiments and previous thermodynamic evaluations. To account for the uncertainty of phase stability, uncertainty quantification based on Bayes' theorem \cite{1763Bayes} has been conducted for calibrating the standard deviations of (CALPHAD) model parameters. Deterministic (metastable) phase diagram and stochastic phase stability were then evaluated via linear-constraint energetic minimizations. Details of this work are hereby reported.

\section*{Computational details}

\noindent \textbf{\large{First-principles calculations.}} \quad In order to evaluate the pseudo-binary phase diagram of Ti$_2$AlC-Cr$_2$AlC, finite-temperature free energies of the MAX and competing phases are needed. For this, first-principles calculations were firstly conducted using the supercell approach, taking into account both vibrational and electronic contributions to the total free energy of each system. Here, the vibrational contribution were evaluated using the quasi-harmonic supercell approach \cite{vanderwalle}. In particular, 6 volumes equally ranging from -2\% to 3\% of the equilibrium volume were considered. For each of these volumes, supercells were constructed and atomic positions were distorted away from equilibrium. First-principles calculations were then conducted to evaluated the atomic forces required to relax the distorted atoms back to their equilibrium position.

For each of these first-principles calculations, the following details hold. The calculation was performed within the framework of DFT \cite{ra-kohnsham}, as implemented in the Vienna ab-initio simulation package (VASP) \cite{ra-vasp,ra-vasp2}. The general gradient approximation (GGA) in the form of PBE \cite{1996Perdew} was employed for the exchange--correlation energy, in conjunction with the projector augmented-wave (PAW) pseudo-potentials formalism \cite{ra-PAW} with $p$ semi-core states treated as valance states. The Brillouin zone integrations were performed using a Monkhorst-Pack mesh \cite{ra-monkpack} with 3000 $k$-points per reciprocal atom. Full relaxations were realized by using the Methfessel-Paxton smearing method of order one \cite{ra-smering} and a final self-consistent static calculation with the tetrahedron smearing method along with Bl{\"o}chl corrections \cite{ra-tetramethod}. A cutoff energy equivalent to 1.3 maximum cut-off energy among the constituent elements was set for each calculation and spin polarization was taken into account.

The calculated forces, required to relax the distorted atoms back to their equilibrium position, were then used to evaluate the dynamical matrix which in turn yields the phonon density of the system at each volume. From here, vibrational enthalpy and entropy as functions of temperature can be derived and finite-temperature free energy can be achieved \cite{vanderwalle}:

\begin{equation}
F_{vib}(T)=k_BT\int_0^{\infty} \ln \left[ 2 \sinh \left( \frac{h\nu}{2k_BT} \right) \right] g(\nu) \ d\nu
\label{eq:VibE}
\end{equation}
in which $k_B$ is Boltzmann's constant, $h$ is Planck's constant, $T$ is temperature, and $g(\nu)$ is the phonon DOS of the structure at equilibrium.

The electronic contribution to the vibrational energy can be readily evaluated as follows \etal \cite{Asta}:

\begin{equation}
F_{el}(T)=E_{el}(T)-TS_{el}(T)
\label{eq:EleE}
\end{equation}
\begin{equation}
E_{el}(T)=\int n(\epsilon) f \epsilon d \epsilon - \int^{\epsilon _F} n(\epsilon) \epsilon \ d \epsilon
\end{equation}
\begin{equation}
S_{el}(T)=-k_B \int n(\epsilon) \left[ f \ln f + (1-f) \ln (1-f) \right] \ d \epsilon
\end{equation}
where, $f$ is the Fermi distribution function and $n(\epsilon)$ is the electronic DOS corresponding to each quasi-harmonic volume at each energy $\epsilon$.

From both vibrational and electronic contributions, the total free energy of the system at each quasi-harmonic volume can then be evaluated:

\begin{equation}
F_{Total}(T)=E_{0K}+F_{vib}(T)+F_{el}(T)
\label{eq:TotalE}
\end{equation}
where, $E_{0K}$ is the ground-state equilibrium energy. Putting all $F_{Total}$ together, an energy surface, $F(V,T)$, can be constructed. The finite-temperature free energy of the system can then be derived by evaluating the equilibrium (minimum) energy $F_0$ of the $F-V$ equation of state at each temperature $T$. It has been shown from previous literature that the supercell approach to calculate finite-temperature energy can yield acceptable results for MAX phases \cite{duong2011, 2014Thomas}.

Free energies were calculated for MAX solid solution and competing phases. Here, the MAX solid solution was modeled using 32-atom special quasi-random structures. Considered compositions include Cr$_{6.25}$, Cr$_{12.5}$,  Cr$_{18.75}$, Cr$_{25}$, Cr$_{50}$, Cr$_{75}$, and Cr$_{87.5}$. Pre-estimations of phonon frequencies at the ground-state equilibrium volumes demonstrated that the solid solutions are mechanically stable up to $12.5$ $at.\%$ Ti in (Cr,Ti)$_2$AlC. Upon examining the $18.75$ $at.\%$ Ti composition and above, we unfortunately experienced ill-posed dynamical matrices which prevent the evaluations of the (Cr,Ti)$_2$AlC's phonon frequencies using the ATAT package. By conducting rough estimations of the dynamical matrices using a few perturbation configurations, we observed that there exist negative frequencies. This tends to indicate that the alloys are likely to be unstable above $18.25$ $at.\%$ Ti under low temperature conditions, and as such, may explain the difficulty in estimating their finite-free energies.

\begin{table*}[ht]
	\centering
	\caption{List of considered competing unary, binary, ternary, and quaternary intermetallic compounds}
	\begin{tabular}{|ccccccc|}
		\hline
		\rowcolor{LightCyan}
		\multicolumn{7}{|c|}{\textbf{Unary}} \\ \hline
		\multicolumn{1}{|c}{\textbf{M$^1$}} & & \multicolumn{1}{c}{\textbf{M$^2$}} & & \multicolumn{1}{c}{\textbf{A}} & & \multicolumn{1}{c|}{\textbf{X}} \\
		\multicolumn{1}{|c}{Ti} & & \multicolumn{1}{c}{Cr} & & \multicolumn{1}{c}{Al}  & & \multicolumn{1}{c|}{C}  \\ \hline
		\rowcolor{LightCyan}
		\multicolumn{7}{|c|}{\textbf{Binary}} \\ \hline
		\textbf{M$^1$-M$^2$} & & \textbf{M$^{1,2}$-A} & & \textbf{M$^{1,2}$-X} & & \textbf{A-X} \\
		\multirow{2}{*}{TiCr$_2$} & & TiAl, Ti$_3$Al, TiAl$_2$, TiAl$_3$ & & TiC, & & \multirow{2}{*}{Al$_4$C$_3$} \\
		& & Cr$_2$Al, Cr$_3$Al, Cr$_5$Al$_8$, $Cr_4Al_9$, $CrAl_4$, $Cr_2Al_{11}$, $Cr_2Al_{13}$ & & Cr$_3$C$_2$, $Cr_7C_3$ Cr$_{23}$C$_6$ & & \\ \hline
		\rowcolor{LightCyan}
		\multicolumn{7}{|c|}{\textbf{Ternary}} \\ \hline
		\multicolumn{7}{|c|}{Cr$_2$AlC, Ti$_2$AlC, Ti$_3$AlC, Ti$_3$AlC$_2$, Ti$_4$AlC$_3$} \\ \hline
		\rowcolor{LightCyan}
		\multicolumn{7}{|c|}{\textbf{Quaternary}} \\ \hline
		\multicolumn{7}{|c|}{(Ti$_{1/2}$Cr$_{1/2}$)$_2$AlC, (Cr$_{2/3}$Ti$_{1/3}$)$_3$AlC$_2$} \\ \hline
	\end{tabular}
	\label{tab:CompetingPhases}
\end{table*}

Competing phases consist of the MAX end-members and other intermetallic compounds that have been observed or proposed in previous literature on the phase stability of Ti$_2$AlC-Cr$_2$AlC. In addition, we also considered the compounds from the assessed phase diagrams of Ti-Cr \cite{1978aKaufman}, Ti-Al \cite{2012Wang}, Ti-C \cite{1978bKaufman}, Cr-Al \cite{2011Wang}, Cr-C \cite{1978bKaufman}, and Al-C \cite{1978bKaufman}. \tab{tab:CompetingPhases} summarizes the competing compounds that we consider in the current work. It should be noted here that within this table, we have excluded from our first-principles calculations Cr$_7$C$_3$, since the dynamical matrix of this phase is close to singularity and the numerical estimation of its eigen values, which are required to evaluate the phonon frequencies, was not possible (similar to the case of the solid solutions above). Other excluded phases are Al$_4$Cr, Al$_9$Cr$_5$, Al$_{11}$Cr$_2$, and Al$_{13}$Cr$_2$ due either to the lack of crystallographic information or expensive, large unit cells. Also, off-stochiometry is not considered within the scope of this work. Such off-stochiometry, even though required for a comprehensive phase-equilibria estimation, generally does not tend to affect the topology of the phase diagram in a significant manner. The ordered $(Ti_{1/2}Cr_{1/2})_2AlC$ structure is found from our previous high-throughput cluster expansion study \cite{2016Talapatra}. It has an alternate order of Ti and Cr in the M layers. The calculated free energies of competing intermetallic compounds and solid solution are tabulated at discrete temperatures in \tab{tab:FreeEnergies}. 


\begin{table*}[htbp]
	\centering
	\caption{Calculated free energies of the considered intermetallic compounds, tabulated at discrete temperatures. Here, the used potentials are PAW-PBE with the $p$ semi-core states treated as valence states. }
	\begin{tabular}{rrrrrrrr}
		\hline
		\rowcolor{LightCyan}
		\multicolumn{1}{c}{\textbf{Composition}} & \multicolumn{7}{c}{\textbf{Energy (eV/f.u.)}} \\
		\hline
		\multirow{2}{*}{\textbf{Unary}} & \multicolumn{1}{c}{\textbf{T = 0 K}} & \multicolumn{1}{c}{\textbf{T = 250 K}} & \multicolumn{1}{c}{\textbf{T = 500 K}} & \multicolumn{1}{c}{\textbf{T = 750 K}} & \multicolumn{1}{c}{\textbf{T = 1000 K}} & \multicolumn{1}{c}{\textbf{T = 1250 K}} & \multicolumn{1}{c}{\textbf{T = 1500 K}} \\\cline{2-8}
		&       &       &       &       &       &       &  \\
		\textbf{M} &       &       &       &       &       &       &  \\
		Cr    & -3.742 & -3.771 & -3.861 & -3.987 & -4.138 & -4.311 & -4.502 \\
		Ti    & -7.897 & -7.929 & -8.024 & -8.154 & -8.308 & -8.482 & -8.672 \\
		\textbf{A} &       &       &       &       &       &       &  \\
		Al    & -3.742 & -3.771 & -3.861 & -3.987 & -4.138 & -4.311 & -4.502 \\
		\textbf{X} &       &       &       &       &       &       &  \\
		C     & -9.221 & -9.225 & -9.245 & -9.284 & -9.339 & -9.409 & -9.490 \\
		\multirow{2}{*}{\textbf{Binary}} &       &       &       &       &       &       &  \\\cline{2-8}
		&       &       &       &       &       &       &  \\
		\textbf{M-M} &       &       &       &       &       &       &  \\
		TiCr$_2$ & -27.513 & -27.580 & -27.810 & -28.141 & -28.542 & -28.997 & -29.498 \\
		\textbf{M-A} &       &       &       &       &       &       &  \\
		AlCr$_2$ & -23.401 & -23.463 & -23.685 & -24.009 & -24.404 & -24.853 & -25.349 \\
		AlCr$_3$ & -32.755 & -32.844 & -33.151 & -33.591 & -34.123 & -34.726 & -35.389 \\
		Al$_8$Cr$_5$ & -78.898 & -79.249 & -80.327 & -81.844 & -83.664 & -85.718 & -87.964 \\
		TiAl  & -12.446 & -12.493 & -12.653 & -12.883 & -13.160 & -13.475 & -13.820 \\
		Ti$_3$Al & -28.538 & -28.639 & -28.969 & -29.437 & -30.001 & -30.640 & -31.344 \\
		TiAl$_2$ & -16.672 & -16.735 & -16.959 & -17.283 & -17.678 & -18.128 & -18.623 \\
		TiAl$_3$ & -20.712 & -20.797 & -21.097 & -21.533 & -22.063 & -22.666 & -23.329 \\
		\textbf{M-X} &       &       &       &       &       &       &  \\
		Cr$_3$C$_2$ & -47.911 & -47.977 & -48.249 & -48.678 & -49.221 & -49.853 & -50.561 \\
		Cr$_{23}$C$_6$ & -279.385 & -279.908 & -281.839 & -284.725 & -288.292 & -292.394 & -296.947 \\
		TiC   & -18.738 & -18.757 & -18.848 & -18.997 & -19.190 & -19.416 & -19.671 \\
		\textbf{A-X} &       &       &       &       &       &       &  \\
		Al$_4$C$_3$ & -43.320 & -43.397 & -43.724 & -44.252 & -44.927 & -45.718 & -46.604 \\
		\multirow{2}{*}{\textbf{Ternary}} &       &       &       &       &       &       &  \\\cline{2-8}
		&       &       &       &       &       &       &  \\
		\textbf{M-M-A} &       &       &       &       &       &       &  \\
		TiAlCr$_2$ & -30.911 & -31.024 & -31.376 & -31.868 & -32.456 & -33.119 & -33.843 \\
		Ti$_2$AlCr & -29.906 & -30.016 & -30.359 & -30.840 & -31.417 & -32.069 & -32.783 \\
		TiAl$_2$Cr & -25.609 & -25.752 & -26.124 & -26.632 & -27.235 & -27.912 & -28.652 \\
		\textbf{M-M-X} &       &       &       &       &       &       &  \\
		Ti$_2$AlC & -31.563 & -31.630 & -31.880 & -32.255 & -32.717 & -33.247 & -33.832 \\
		Cr$_2$AlC & -32.959 & -33.018 & -33.249 & -33.602 & -34.043 & -34.552 & -35.116 \\
		Ti$_3$AlC & -39.612 & -39.723 & -40.086 & -40.610 & -41.249 & -41.977 & -42.779 \\
		Ti$_3$AlC$_2$ & -50.463 & -50.548 & -50.885 & -51.406 & -52.057 & -52.811 & -53.649 \\
		Ti$_4$AlC$_3$ & -69.204 & -69.306 & -69.732 & -70.400 & -71.243 & -72.223 & -73.316 \\
		\multirow{2}{*}{\textbf{Quaternary}} &       &       &       &       &       &       &  \\\cline{2-8}
		&       &       &       &       &       &       &  \\
		(Ti$_{1/2}$,Cr$_{1/2}$)$_2$AlC & -32.325 & -32.388 & -32.628 & -32.993 & -33.450 & -33.978 & -34.565 \\
		(Cr$_{2/3}$,Ti$_{1/3}$)$_3$AlC$_2$ & -51.716 & -51.800 & -52.136 & -52.659 & -53.317 & -54.081 & -54.933 \\
		&       &       &       &       &       &       & \\
		\multirow{2}{*}{\textbf{Solid solution}} &       &       &       &       &       &       &  \\\cline{2-8}
		&       &       &       &       &       &       &  \\
		(Ti$_{12.5}$,Cr$_{87.5}$)$_2$AlC & -32.673 & -32.734 & -32.972 & -33.336 & -33.792 & -34.319 & -34.906 \\
		&       &       &       &       &       &       & \\		
		\bottomrule
	\end{tabular}%
	\label{tab:FreeEnergies}%
\end{table*}%

\vspace{2.ex} \noindent \textbf{\large{CALPHAD.}} \quad Although first-principles calculation is advantageous in sense that reasonable thermodynamic properties can be derived based on simple knowledge of possible existing phases and their crystal structure, such an approach apparently does have limitations, for instance the technical issues involving phonon frequencies or crystal structure information, as aforementioned. Due to this, comprehensive knowledge on phase stability may not be achieved even with powerful high-throughput computational facilities. To compensate for this lack in capability of first-principles calculations, CALPHAD methodology was integrated into our computational framework. Here, the use of CALPHAD, in coupling with first-principles calculations, would introduce additional high-temperature phase-equilibria constraints that allow the calculated Gibbs free energies to be modified in such a way that they satisfy both quantum-mechanical-based calculations and experimental phase equilibria. This way, the assessed thermodynamics of the systems are guaranteed high consistency and are therefore more reliable. Another advantage of the coupling approach is that missing phases from first-principles calculations, such as Cr$_7$C$_3$, Al$_4$Cr and Al$_9$Cr$_5$, can be taken into account via their relative thermodynamic relationships with the assessable phases.

\begin{table*}[htbp]
	\centering
	\caption{Optimized CALPHAD parameters and their (Bayesian) standard deviations of the binaries, ternaries, and quaternary. Note that (1) the energies resulted from these parameters are in the typical unit of $J/mol/atom$ and (2) the standard deviations of Cr-Al-C and Ti-Al-C are relatively large due to the fact that there are much fewer thermodynamic constraints available for these systems than the binaries.}
	\resizebox{\textwidth}{!}{%
	\begin{tabular}{lrllr}
		\hline
		\rowcolor{LightCyan} \multicolumn{2}{c}{Ti-Al} & & \multicolumn{2}{c}{Cr-Al} \\
		\hline
		& & & & \\
		Liquid & -44826.774 ($\pm$5940.548) - 9.705 ($\pm$3.425) $\times$ T & & Liquid & -16280.734 ($\pm$2547.996) - 1.255 ($\pm$0.00581) $\times$ T  \\
		& 19984.519 ($\pm$6532.623) - 3.768 ($\pm$4.379) $\times$ T & & Bcc & -31542.024 ($\pm$4864.770) + 3.55 ($\pm$2.942) $\times$ T \\
		Fcc & GHSERAL &       & & 21170.075 ($\pm$11420.795) -10.532 ($\pm$7.273) $\times$ T \\
		Al$_3$Ti & -161609.797 ($\pm$2728.262) + 34.147 ($\pm$1.876) $\times$ T & & Fcc   & 36281.545 ($\pm$2922.821) - 39.185 ($\pm$2.156) $\times$ T  \\
		Al$_2$Ti & -130310.371 ($\pm$2455.017) + 27.823 ($\pm$1.767) $\times$ T & & AlCr$_2$ & -22220.569 ($\pm$3867.136) - 10.415 ($\pm$2.592) $\times$ T  \\
		AlTi & -777301.828 ($\pm$10896.578) + 100.0 ($\pm$7.655) $\times$ T & & Al$_4$Cr & -33435.227 ($\pm$3432.087) - 11.231 ($\pm$2.627) $\times$ T  \\
		AlTi$_3$ & -256328.239 ($\pm$9668.292) + 14.891 ($\pm$6.253) $\times$ T & & Al$_8$Cr$_5^L$ & -34476.155 ($\pm$24897.260) - 99.999 ($\pm$14.365) $\times$ T  \\
		Hcp & -85725.167 ($\pm$1232.210) + 0.149 ($\pm$0.236) $\times$ T & & Al$_8$Cr$_5^H$ & -80955.010 ($\pm$15349.512) - 67.728 ($\pm$11.668) $\times$ T  \\
		& -3664.848 ($\pm$4092.983) + 4.421 ($\pm$2.452) $\times$ T & & Al$_9$Cr$_4^L$ & -21193.384 ($\pm$20765.556) - 99.984 ($\pm$14.385) $\times$ T  \\
		& -18016.218 ($\pm$7117.0) + 27.146 ($\pm$4.157) $\times$ T & & Al$_9$Cr$_4^H$ & -83347.404 ($\pm$13272.196) - 52.337 ($\pm$9.686) $\times$ T  \\
		Bcc & -77446.353 ($\pm$2772.679) - 1.560 ($\pm$1.607) $\times$ T & & Al$_{11}$Cr$_2$ & -89989.489 ($\pm$6663.153) - 12.427 ($\pm$5.063) $\times$ T  \\
		& -5001.401 ($\pm$4993.332) + 4.004 ($\pm$3.061) $\times$ T & & Al$_{13}$Cr$_2$ & -108076.0 ($\pm$5847.849) + 2.893 ($\pm$3.921) $\times$ T  \\
		& 23443.277 ($\pm$8667.487) - 5.158 ($\pm$4.838) $\times$ T &  &     &  \\
		Al$_{11}$Ti$_5$ & -617287.480 ($\pm$17072.163) + 100.0 ($\pm$11.203) $\times$ T & &      &  \\
		& & & & \\
		\hline
		\rowcolor{LightCyan} \multicolumn{2}{c}{Ti-C} & & \multicolumn{2}{c}{Ti-Cr} \\
		\hline
		& & & & \\
		Liquid & -29519.275 ($\pm$13942.121) -70.379 ($\pm$5.40) $\times$ T & & Liquid & 25474.148 ($\pm$7201.615) -19.291 ($\pm$4.464) $\times$ T \\
		Bcc & GBCCTI &      & & 9474.205 ($\pm$7013.502) -4.579 ($\pm$4.299) $\times$ T \\
		Hcp & GHSERTI & & Bcc   & -3127.928 ($\pm$1738.686) +5.743 ($\pm$1.636) $\times$ T \\
		TiC & -1410506.470 ($\pm$55337.817) -37.795 ($\pm$21.571) $\times$ T &      & & 5016.857 ($\pm$1187.926) -3.643 ($\pm$1.107) $\times$ T \\
		Graphite & GHSERCC &      & & 1294.061 ($\pm$1386.114) -3.673 ($\pm$1.004) $\times$ T \\
		&      & & TiCr$_2$ & -37098.412 ($\pm$900.327) +8.455 ($\pm$0.790) $\times$ T \\
		&      & & Hcp   & 23631.773 ($\pm$1522.291) -4.893 ($\pm$2.040) $\times$ T \\
		& & & & \\
		\hline
		\rowcolor{LightCyan} \multicolumn{2}{c}{Cr-C} & & \multicolumn{2}{c}{Al-C} \\
		\hline
		& & & \\
		Liquid & -127800.114 ($\pm$682.547) + 0.0 ($\pm$0.0571) $\times$ T & & Liquid & 109830.637 ($\pm$5334.934) - 59.436 ($\pm$2.461) $\times$ T \\
		Bcc & GHSERCR & & Fcc   & GHSERAL \\
		Cr$_{23}$C$_6$ & -229313.680 ($\pm$9578.358) -54.894 ($\pm$5.772) $\times$ T & & Al$_4$C$_3$ & -22342.766 ($\pm$13822.148) - 24.358 ($\pm$6.117) $\times$ T \\
		Cr$_7$C$_3$ & -60777.111 ($\pm$4392.626) -48.386 ($\pm$2.355) $\times$ T & & Graphite & GHSERCC \\
		Cr$_3$C$_2$ & -24866.751 ($\pm$2832.234) -24.788 ($\pm$1.457) $\times$ T &   &   &  \\
		& & & & \\
		\hline
		\rowcolor{LightCyan} \multicolumn{2}{c}{Cr-Al-C} & & \multicolumn{2}{c}{Ti-Al-C} \\
		\hline
		& & & & \\
		Cr$_2$AlC & -73776.041 ($\pm$50610.801) + 13.877 ($\pm$23.171) $\times$ T & & Ti$_2$AlC & -274945.380 ($\pm$337844.323) + 23.478 ($\pm$40.730) $\times$ T \\
		&      & & Ti$_3$AlC & -286333.291 ($\pm$353070.956) + 14.647 ($\pm$25.321) $\times$ T \\
		&      & & Ti$_4$AlC$_3$ & -286333.291 + 14.647 $\times$ T \\
		& & & & \\
		\hline
		\rowcolor{LightCyan} \multicolumn{5}{c}{Ti-Cr-Al-C} \\
		\hline
		& & & & \\
		(Ti$_{1/3}$Cr$_{2/3}$)$_{3}$AlC$_2$ & -286333.291 + 14.647 $\times$ T & & (Ti$_{1/2}$Cr$_{1/2}$)$_2$AlC & -286333.291 + 14.647 $\times$ T \\		
		& & & & \\
		\bottomrule
	\end{tabular}%
	}
	\label{tab:BayesCALP}%
\end{table*}%

\begin{figure}[htbp]
	\centering
	\includegraphics[scale=0.6]{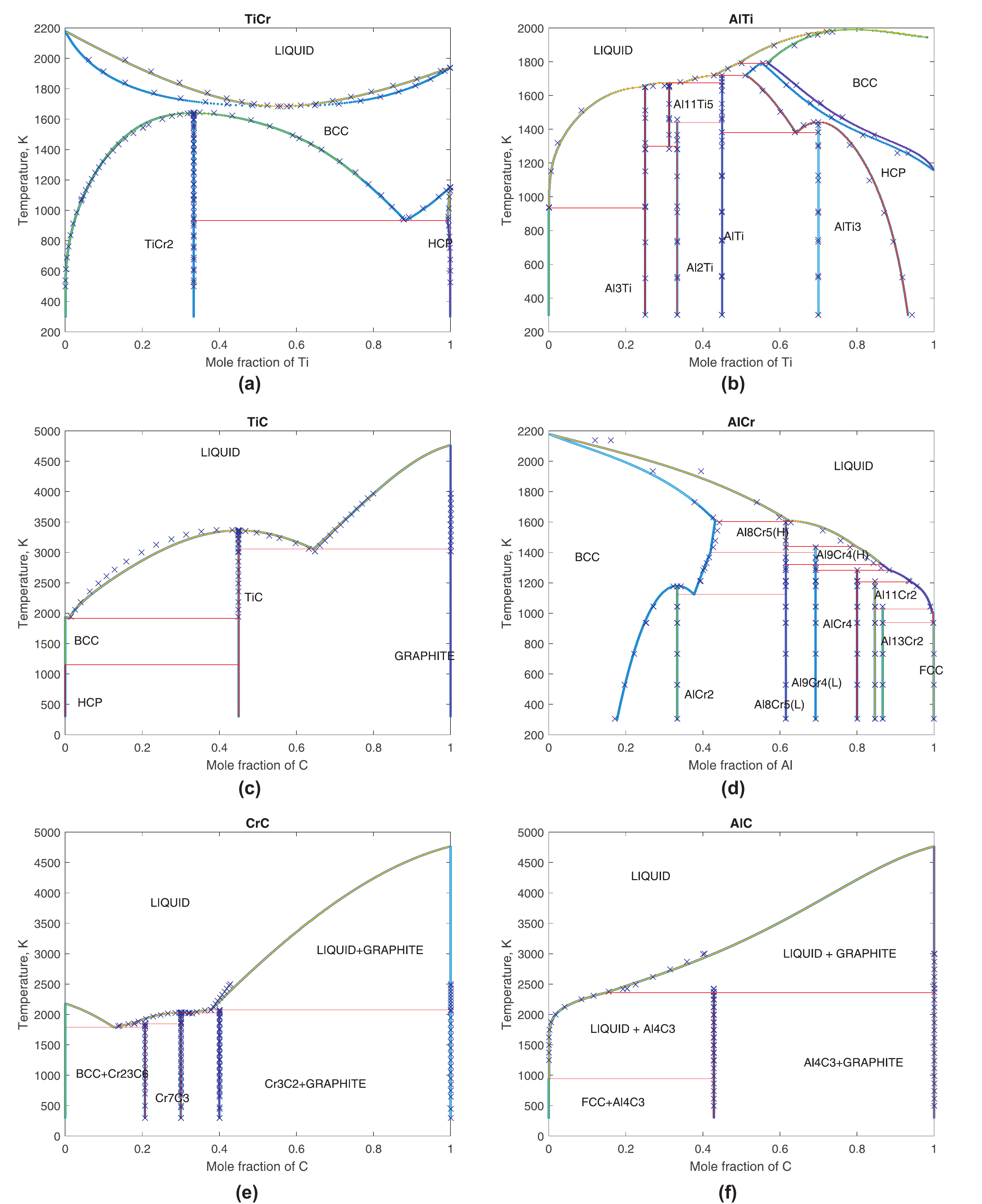}
	\caption[]{Calculated phase diagram of the binary systems. Here, the (blue) cross indicates phase equilibria data derived from the TCFE7 database, with off-stoichimetric data reduced to stoichiometric for the sake of simplicity.}
	\label{fig:BinaryPDs}
\end{figure}

In this work, first-principles-guided CALPHAD assessments have been conducted for six binary systems, including Ti-Cr, Ti-Al, Ti-C, Cr-Al, Cr-C, and Al-C. The changes of energetic references from first-principles to CALPHAD scales were needed for thermochemistry data. This was made possible by the use of the SGTE database \cite{1991Dinsdale} (note that DFT and SGTE energies almost only differ from each other by a constant). Phase equilibria data were calibrated from the TCFE7 database within the Thermo-Calc package \cite{2002Andersson}. For simplicity, the intermetallic compounds were again treated as stoichiometric. Optimized parameters, describing the Gibbs free energies of competing binaries, were reported in \tab{tab:BayesCALP}. The phase diagrams calculated using these parameters are shown in \fig{fig:BinaryPDs}.

For the ternary phases, CALPHAD free energies were obtained by fitting compound-energy model \cite{2001Hillert} to the first-principles data, taking into account experimental phase-equilibria constraints when available. The compound-energy model in this work takes the simple form of:

\begin{equation}
G^{mix}_{M_2AlC} = \frac{a + bT + 2G_{M} + G_{Al} + G_{C}}{4}
\end{equation}
where, $T$ is temperature ($K$), $a$ and $b$ account for the interactions between M, Al, and C elements and the elemental Gibbs energies were taken from the SGTE database. Here, the phase-equilibria data for fitting process are available for Cr$_2$AlC, Ti$_2$AlC, and Ti$_3$AlC from literatures \cite{1980Schuster, 2006Hallstedt, 1994Pietzka} and allow Bayesian estimations of $a$'s and $b$'s standard deviations using the MCMC method, which will be mentioned shortly after. For Ti$_4$AlC$_3$, (Ti$_{1/2}$Cr$_{1/2}$)$_2$AlC, and (Ti$_{1/3}$Cr$_{2/3}$)$_3$AlC$_2$, it is only possible to fit the CALPHAD energies to the first-principles data; and, since there are no fitting errors, standard deviations are unavailable for these phases. The remaining ternaries, contributed from the Ti-Cr-Al system, are not considered within the scope of this work. The reason for this is that these phases are not observed during the synthesis of (Cr,Ti)$_2$AlC and hence are not likely stable. The assessed interaction parameters of the ternary phases are reported in \tab{tab:BayesCALP}.

For the solid solution phase, using the first-principles data limited to 12.5 at.\% Ti alloy, we have attempted to assess a comprehensive description using subregular solution model. In particular, given that $G^o_{Ti_2AlC}(T)$ and $G^o_{Cr_2AlC}(T)$ are the free energies of the end-members, the mixing energy of solid solution, in unit of $J/mol/f.u.$, is expressed as:

\begin{eqnarray}
G^{mix}_{(Ti_xCr_{1-x})_2AlC}(x,T) &=& xG_{Ti_2AlC}(T) + (1-x)G_{Cr_2AlC}(T) + G^{ex}_{hcp-(Cr,Ti)} \nonumber \\
                                   & & + 2RT(x\ln(x)+(1-x)\ln(1-x)) + x(1-x)(a + bT + cT\ln(T)) + \epsilon
\end{eqnarray} 
where, $R$ is the gas constant, the factor of $2$ before $RT$ accounts for the two M-sublattices in the M$_2$AX formula unit, $a$, $b$, and $c$ are $-28112.74$, $303.09$, and $-41.14$, respectively, $G^{ex}_{hcp-(Cr,Ti)}$ is the excess Gibbs energy of mixing in $hcp$-(Cr,Ti) (see \tab{tab:BayesCALP}), and $\epsilon = 13144.380$ is the standard deviations of fitting errors. At temperatures higher than $\sim1710$ $K$, using the energy minimization process (to be mentioned later) we observed the phase segregation of Cr$_2$AlC into Al$_8$Cr$_5$, Al$_4$C$_3$ and Cr$_3$C$_2$. This is found similar to the change of lattice stability \cite{1959Kaufman} in conventional binary system; as such, the mixing energy of solid solution above this temperature is written in term of the segregation products instead of Cr$_2$AlC.

\vspace{2.ex} \noindent \textbf{\large{Bayesian uncertainty quantification.}} \quad To account for the errors of assessed energies and their propagations to phase equilibria of the system, Bayesian quantifications of model uncertainty were implemented. For this, prior and likelihood were assumed to be uniform and Gaussian distributions, respectively. The range of the uniform distribution was from -300\% to +300\% of the assessed parameters and the variance of the Gaussian distribution was initially chosen to be 0.01. Markov Chain Monte Carlo (MCMC) simulation was then conducted to sample 100,000 promising parameter candidates for each binary system and ternary phases, using the Metropolis-Hastings ratio as the selection criteria for sampling the parameter space. In particular, a parameter was randomly generated at each MCMC iteration and would be selected if it had the joint probability of prior and likelihood either higher than that of the current accepted parameter or a random value. During this process, the variance of likelihood was dynamically updated to better describe its corresponding distribution. Upon collecting 100,000 samples, the variance-covariance matrix was estimated (via the Monte-Carlo integration scheme) and the standard deviations of the assessed parameters were derived. The results are shown in \tab{tab:BayesCALP} together with the assessed parameters. For more details of the uncertainty quantification, the readers are referred to the work of Duong \etal \cite{2016Duong}.

Here, the quantification of model uncertainty introduces a different aspect which, from our perspectives, essentially complements the conventional deterministic view of phase equilibria (via the CALPHAD approach) in a stochastic manner. It is interesting to note that this uncertainty quantification integrates naturally with the CALPHAD method just as the CALPHAD method integrates with first-principles calculations. Together, these approaches form a strong integrated computational scheme that can allow satisfactory thermodynamic evaluations and beyond -- as this framework is malleable when there are new experimental and/or theoretical data.

\vspace{2.ex} \noindent \textbf{\large{Energy minimization.}} \quad Based on the energies and their uncertainties evaluated via the first-principles-driven Bayesian CALPHAD scheme above, both deterministic and stochastic phase competitions among the competing phases can the be investigated by means of Gibbs energy minimization with respect to mass-conservation constraints. For this, we have adopted the linear optimization procedure proposed by Sun \etal \cite{2003Sun}. In particular, given that $a^{Ti}$, $a^{Cr}$, $a^{Al}$, and $a^{C}$ are the elemental compositions of Ti, Cr, Al, and C respectively, the linear minimization reads:

\begin{equation}
min\left(E_{comp}(a^{Ti}, a^{Cr}, a^{Al}, a^{C})\right) = min\left(\sum_{i}^{n}x_{i}E_{i}\right)
\end{equation}
where, $x_i$ and $E_i$ are the amount and energy (per formula unit) of compound $i$, and $E_{comp}$ is the energy of composite that contains the most competitive phases. The minimization procedure is subject to the constraints:

\begin{equation}
x_i \ge 0; \quad \sum_{i}^{n}x_{i}^{Ti} = a^{Ti}; \quad \sum_{i}^{n}x_{i}^{Cr} = a^{Cr}; \quad \sum_{i}^{n}x_{i}^{Al} = a^{Al}; \quad \sum_{i}^{n}x_{i}^{C} = a^{C}
\end{equation}

The result of phase competition among the intermetallic compounds via energy minimization at different temperature and composition conditions was then compared against the solution, and the outcomes of this shaped up the pseudo-binary phase diagram of Ti$_2$AlC-Cr$_2$AlC. Here, it should be noted that by separating the solid solution out of the energy minimization process for a second-state phase competition, we more or less favored the stability of solid solution over intermetallic compounds. It also helps to simplify the implementation of the minimization procedure and at the same time increases the procedure's (numerical) precision. We found that this two-stage practice was essential for sketching out an initial phase diagram that could then serve as a reference for future investigations and refinements. In other words, we take, in the current work, the practical view of Integrated Computational Materials Engineering \cite{2008ICME}, i.e. ``\textit{a less-than-perfect solution may still have high impact.}''

\section*{Results and discussions}

\begin{figure}[htbp]
	\centering
	\includegraphics[scale=1.25]{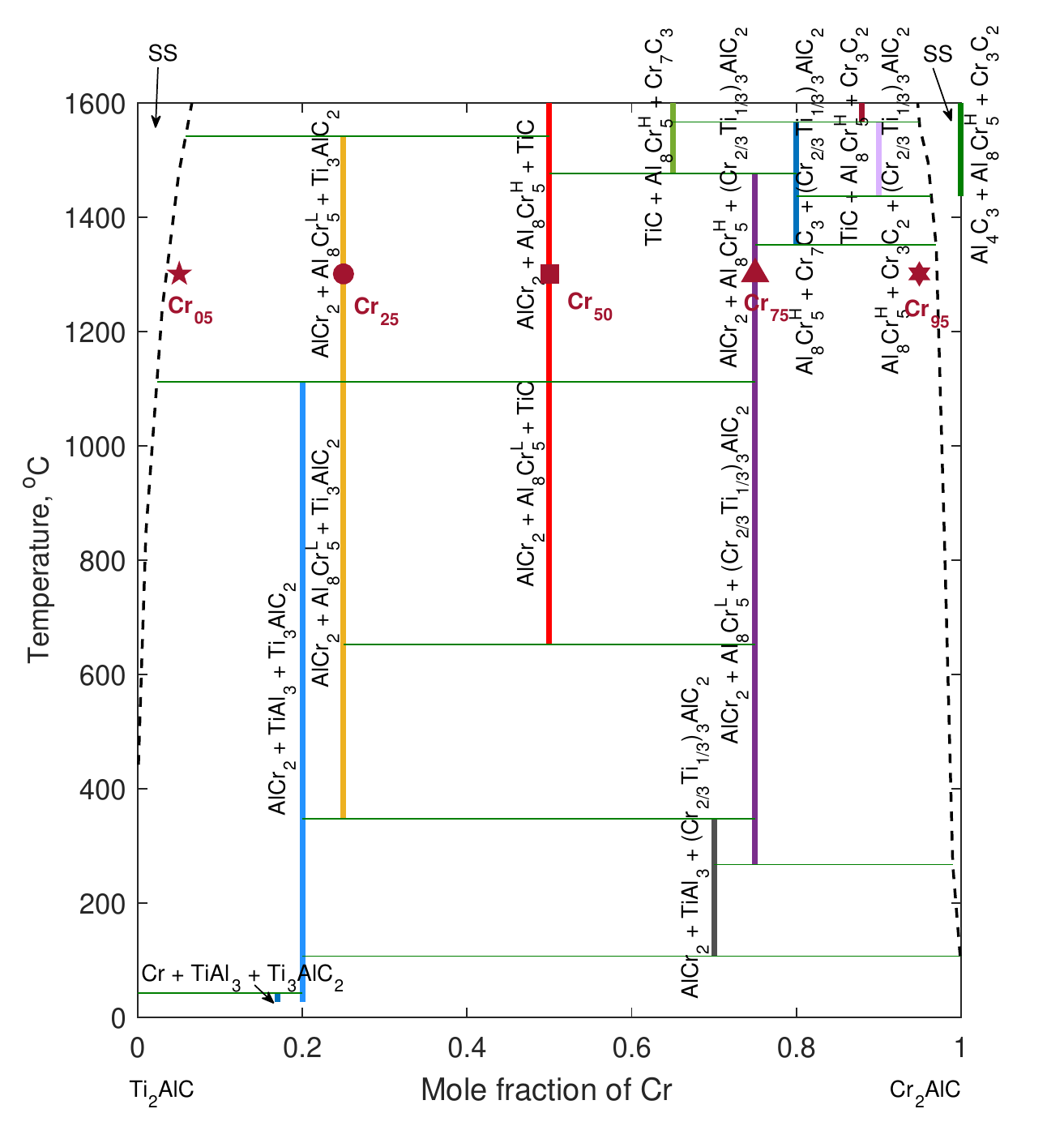}
	\caption[]{Deterministic phase diagram of Ti$_2$AlC-Cr$_2$AlC. Note that above $1410^oC$ Cr$_2$AlC tends to decompose into Al$_4$C$_3$, Al$_8$Cr$_5$, and Cr$_3$C$_2$, as according to the current deterministic evaluation. Here, the symbols indicate experimental data from Horlait \etal \cite{2015Horlait}:
		
		\setlength{\parindent}{4.5cm} \textbf{Cr$_{\textbf{05}}$:} Al$_2$Cr + TiAl$_2$ + (Cr$_{0.02}$Ti$_{0.98}$)$_2$AlC + (Cr$_{0.02}$Ti$_{0.98}$)$_3$AlC$_2$
		
		\setlength{\parindent}{4.5cm} \textbf{Cr$_{\textbf{25}}$:} AlCr$_2$ + Al$_8$Cr$_5$ + TiC + (Cr$_{0.02}$Ti$_{0.98}$)$_3$AlC$_2$
		
		\setlength{\parindent}{4.5cm} \textbf{Cr$_{\textbf{50}}$:} AlCr$_2$ + Al$_8$Cr$_5$ + TiC + (Cr$_{0.95}$Ti$_{0.05}$)$_2$AlC
		
		\setlength{\parindent}{4.5cm} \textbf{Cr$_{\textbf{75}}$:} AlCr$_2$ + Al$_8$Cr$_5$ + TiC + (Cr$_{2/3}$Ti$_{1/3}$)$_3$AlC$_2$
		 
	    \setlength{\parindent}{4.5cm} \textbf{Cr$_{\textbf{95}}$:} Al$_{80}$Cr$_{20}$ + TiC + (Cr$_{0.95}$Ti$_{0.05}$)$_2$AlC
	    
	 }
	\label{fig:MetastablePD}
\end{figure}

\noindent \textbf{\large{Deterministic phase diagram.}} \quad The result of the deterministic phase-diagram evaluation is shown in \fig{fig:MetastablePD}. Before discussing the results, it is worth noting that firstly the current estimated phase diagram is metastable in the sense that possible solid solutions of unary, binary, and ternary systems constituted from Ti, Cr, Al, and C as well as liquid and gases are not considered. Secondly, since liquid and gas are not considered, we restrict our current interpretation of thermodynamic properties of Ti$_2$AlC-Cr$_2$AlC at temperatures lower than the higher melting temperature between the end members, which is $\sim1600^oC$ near Ti$_2$AlC \cite{1994Pietzka, 2006Hallstedt}. With this in mind, let us discuss the result of the current estimated phase diagram as follows.

From \fig{fig:MetastablePD}, it can be seen that the pseudo-binary system generally consists of intermetallic compounds, which divide the phase diagram into many complicated multi-phase regions and feature a strong ordering tendency within the system. Solid solution, as such, is rather weak and only limited to the solute regions at Ti$_2$AlC and Cr$_2$AlC terminals. In particular, near the Ti$_2$AlC side, the solution of Ti$_2$AlC and Cr$_2$AlC only extends up to $\sim 7$ at.\% Cr at $1600^oC$, after forming at $\sim450^oC$. This is consistent with previous experiments, which generally assume a small solubility within this region. For instance, Horlait \etal \cite{2015Horlait} reported a solubility of only $1-2$ $at.$\% Cr in Ti$_2$AlC at $1300^oC$; in our case, the solubility of $\sim 3.5$ $at.$\% Cr in Ti$_2$AlC is observed. Similarly, near the Cr$_2$AlC terminal, the solubility of Ti in Cr$_2$AlC only reaches about $5.5$ at.\% Ti at $1600^oC$, after forming at $\sim 100^oC$. This projects to the solubility of $\sim 3$ at.\% Ti at $1300^oC$ which is slightly less than those of Lee \etal \cite{2010Lee, 2012Lee} and Horlait \etal \cite{2015Horlait} which are $\sim 5$ $at$\% Ti.

Compared to the recent experiment by Horlait etal \cite{2015Horlait}, interesting agreement and disagreement in term of the system's phase stability can be found. In particular,

\begin{itemize}
	\item At 5 at.\% Cr, Horlait \etal reported the coexistence of (Ti$_{0.98}$Cr$_{0.02}$)$_2$AlC, (Ti$_{0.98}$Cr$_{0.02}$)$_3$AlC$_2$, Al$_2$Cr and TiAl$_2$, while in our case we find (Ti$_{0.982}$Cr$_{0.018}$)$_2$AlC, and Ti$_3$AlC$_2$ -- which may be considered equivalent to the observed off-stoichiometric MAX phase as indeed Horlait \etal has reported that the lattice parameters of this off-stoichiometric phase is not much different from those of Ti$_3$AlC$_2$ -- yet they coexist with the high-temperature Al$_8$Cr$_5$ and AlCr$_2$ phases instead of TiAl$_2$ and Al$_2$Cr. Here, it is believed that the difference between computation and experiment is due to the fact that Al$_2$Cr is not considered as one competing phase in the current evaluation. The reason for this is that Al$_2$Cr is not found among the equilibrium phases within the Al-Cr phase diagram \cite{2011Wang}.
	
	\item At 25 at.\% Cr, Horlait \etal found AlCr$_2$, Al$_8$Cr$_5$, (Ti$_{0.98}$Cr$_{0.02}$)$_3$AlC$_2$, and TiC coexisting with each other. In our case, we also find AlCr$_2$, Al$_8$Cr$_5$ and Ti$_3$AlC$_2$ -- which, again, may be considered as the off-stoichiometric (Ti$_{0.98}$Cr$_{0.02}$)$_3$AlC$_2$ --  but not TiC.
	
	\item At 50 at.\% Cr, the experiment reported a predominant existence of TiC ($\sim42$ vol\%) and (Cr$_{0.95}$Ti$_{0.05}$)$_2$AlC ($\sim21$ vol\%) as well as AlCr$_2$ and Al$_8$Cr$_5$. From our (deterministic) evaluation, there is indeed a predominance of TiC coexisting with AlCr$_2$ and Al$_8$Cr$_5$, but not (Cr$_{0.95}$Ti$_{0.05}$)$_2$AlC.
	
	\item At 75 at.\% Cr, the experiment reported a clear evidence of the existence of (Cr$_{2/3}$Ti$_{1/3}$)$_3$AlC$_2$ which coexists with AlCr$_2$, Al$_8$Cr$_5$, and TiC. In our case, we also observe the ordered 312-MAX phase together with AlCr$_2$ and Al$_8$Cr$_5$. At this composition, we do not find the existence of TiC anymore.
	
	\item At 95 at.\% Cr, the experiment indicated a predominance of (Cr$_{0.95}$Ti$_{0.05}$)$_2$AlC with minor Al$_{80}$Cr$_{20}$ and TiC while in our case we find (Cr$_{0.98}$Ti$_{0.02}$)$_2$AlC solid solution with minor phases as AlCr$_2$, Al$_8$Cr$_3$, and (Cr$_{2/3}$Ti$_{1/3}$)$_3$AlC$_2$. Here, since the crystallography of Al$_{80}$Cr$_{20}$ is unclear and hence can not be confidently related to the equilibrium phases of Al-Cr \cite{2011Wang}, we can not account properly the (relative) stability of this phase. One possible assumption would be that Al$_{80}$Cr$_{20}$ represents Al$_4$Cr, but even Al$_4$Cr is not observed in the current deterministic evaluation.
	
\end{itemize}

\begin{figure}[htbp]
	\centering
	\includegraphics[scale=1.35]{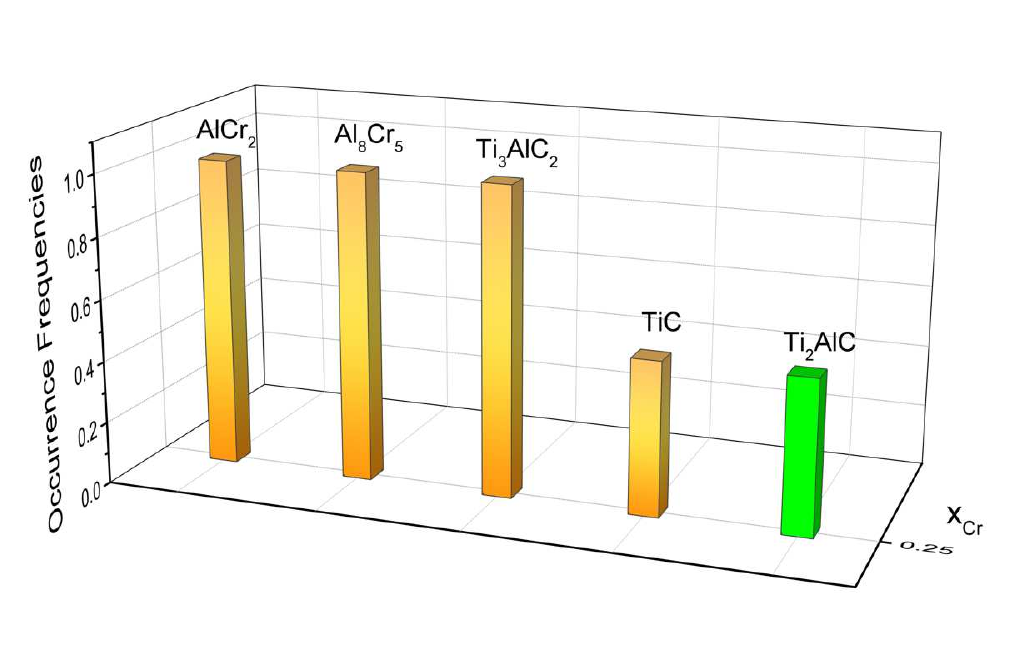}
	\caption[]{Occurrence frequencies of stable phases with Al and C ratios varying between $0.95-1.05$ at $1300^oC$. Here, it can be seen that all experimentally observed phases can be found.}
	\label{fig:PhaseFreq25}
\end{figure}

Here, it should be noted that, except for the cases resulting from the lack of crystallographic information, most of the disagreements with experiments are actually due to the fact that uncertainty is not accounted for within the deterministic evaluation. For instance, at 25 $at.$\% Cr the observation of TiC, according to our current estimation, is indeed possible when the composition of Al or C slightly deviate from the stoichiometric values. This is demonstrated in \fig{fig:PhaseFreq25}, in which the occurrence frequencies of stable phases resulting from 10,000 Gibbs minimization processes with random selections of Al and C ratios in between $0.95-1.05$ have been shown. Given that the starting composition within the experiment was 2:1.05:0.95 (to account for Al sublimation and C addition during the hot pressing synthesis with carbon die), it is plausible that some uncertain effective non-stoichiometric ratio (differing from 2:1:1) is responsible for the experimental products. Such uncertainty although exists here and generally in many other cases is never a natural part of conventional (deterministic) phase-stability evaluations. Considering that equilibrium analysis is computationally-driven and that computation has various sources of ambivalence stemming from the theories and/or experimental data that it relies upon, the real pictures of phase stability could be much different from those for instance shown in \fig{fig:MetastablePD} and \fig{fig:PhaseFreq25} and can only be evaluated when uncertainty is accounted for.

\begin{figure}[htbp]
	\centering
	\includegraphics[scale=0.45]{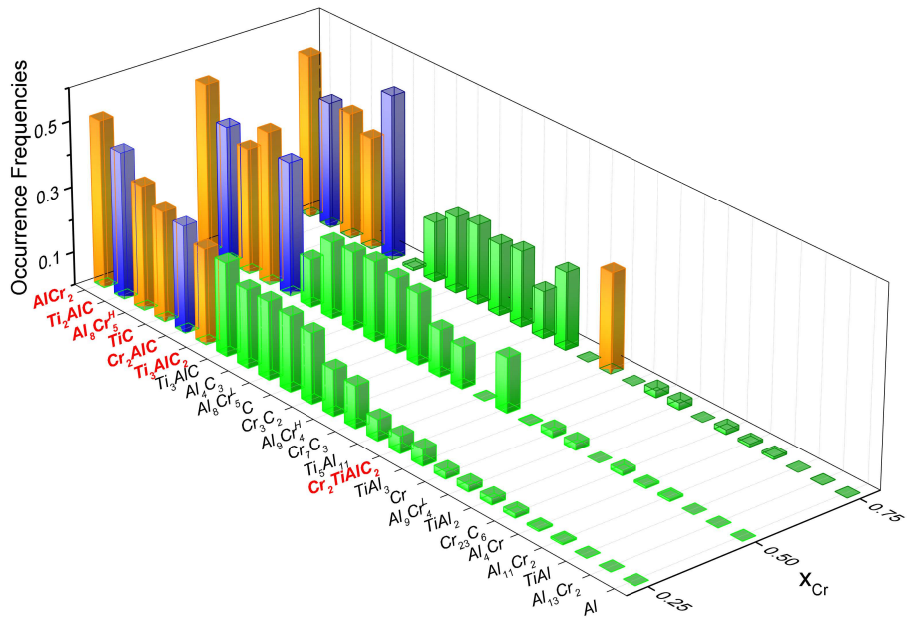}
	\caption[]{Stochastic phase stabilities of $Cr_{25}$, $Cr_{50}$, and $Cr_{75}$ at $1300^oC$. Note that other possible stable phases exist, as the result of high degree of uncertainty introduced via all model parameters of all the considered phases.}
	\label{fig:PhaseDistro}
\end{figure}

\vspace{2.ex} \noindent \textbf{\large{Stochastic phase stability.}} \quad There are different ways to account for the effect of uncertainty on phase-stability analysis. Within the current work, we take the simple approach as follows. Firstly, a large number of values (100,000 in this case) within the derived standard deviations of the assessed model parameters as well as $\pm 5\%$ deviations of Al and C mass ratio were sampled using the latin hypercube algorithm \cite{2008Iman} with minimizing correlation criteria. These parameters were then used as inputs to the Gibbs energy minimization code in order to evaluate the most stable phases at 25, 75, and 95 $at.$\% Cr and at $1300^oC$. Note here that due to the lack of crystallographic information of Al$_2$Cr and Al$_{80}$Cr$_{20}$ at 5 and 95 $at.$\% Cr respectively, phase stability at these compositions can not be fairly assessed, and hence is not considered in the current stochastic analysis. It should be noted here that the missing of Al$_2$Cr and Al$_{80}$Cr$_{20}$ does not tend to affect phase stability at 25, 75, and 95 $at.$\% Cr as they do not seem to be stable phases at these compositions \cite{2014Ying, 2015Horlait}. Also, it should not affect much the solubility of the system near the solute region since it has been shown experimentally\cite{2015Horlait} that Al$_2$Cr and Al$_{80}$Cr$_{20}$ are minor phases. For future evaluations, it is required experimental clarifications on the existences as well as crystal structures of these phases. 
Resulting stable phases from the above Gibbs energy minimizations were then counted and their occurrence frequencies, which intuitively represent the likelihoods of observing the stable phases, were derived.

The results of the stable-phase survey are reported in \fig{fig:PhaseDistro}. From this figure, it can be seen that computation and experiment are generally in good agreement with each other in the sense that the experimentally observed phases (in orange) are the ones that are most likely to occur, or in other words most likely to be stable phases. Here, it is noted that, since the solid solution is not accounted for, its occurrence frequencies may be interpreted based on those of Cr$_2$AlC and Ti$_2$AlC (in blue). In this regards, the stochastic analysis seems reasonable in the sense that it tends to recommend a higher probability of having Ti-rich solid solution near Ti$_2$AlC side (i.e. higher Ti$_2$AlC occurrence frequency than Cr$_2$AlC's) and a higher probability of having Cr-rich solid solution near Cr$_2$AlC side (i.e. higher Cr$_2$AlC occurrence frequency than Ti$_2$AlC's). At the 50 $at.$\% Cr composition, it can be seen from \fig{fig:PhaseDistro} that both Ti$_2$AlC and Cr$_2$AlC have almost the same occurrence frequencies. Intuitively speaking, this is statistically eligible and may perhaps be plausible as the system does not seem to favor one end-member over the other according to their (close) solute mixing behaviors. From here, it may be inferred that the system has a relatively high tendency to form the (Ti$_{50}$Cr$_{50}$)$_2$AlC solution at this composition-temperature condition. Given that the system tends to possess a strong ordering tendency but still demonstrate a small solubility near the two solute regions, the solution here may be interpreted as a metastable state, which simply appears as one of the highly possible phases due to the fact that the more stable state is not present. In this regards, the metastable state may well decompose towards either Ti$_2$AlC or Cr$_2$AlC side but likely preserve some solubility of one within the other (e.g. spinodal decomposition), and perhaps therefore this corresponds to the experimentally observed (Cr$_{0.95}$Ti$_{0.05}$)$_2$AlC. Additional experiments would help to refine better the confidence in the stochastic picture of phase stability among the competing phases in the Ti$_2$AlC-Cr$_2$AlC system.


Other than these, it can be seen from \fig{fig:PhaseDistro} that there are also other competing phases with high occurrence probabilities, relative to the experimentally observed phases. This is not very much surprising considering that there are many model parameters, and hence a high degree of uncertainty, affecting the outcomes of the stochastic analysis. Here, it is interesting to note that the observations of all possible phases feature well the differences between the conventional and current approaches to the phase-stability problem, namely different attacking questions such as ``what are (actually) the stable phases?'' v.s. ``what are likely to be stable phases?'' and their answers, respectively. At first, the results of stochastic phase stability may appear indecisive if not overall confusing but a feature of this analysis is the fact that it provides a comprehensive account for the likely stable and/or metastable phases that one can expect from a synthesis exercise, accounting for uncertainties not only in the models used for the predictions but also in the experiments themselves. Additionally, so long as better and more data is supplied, the malleable integrated approach can be refined so it eventually converges to the deterministic regime with high confidence. This is one important but so-far-missing feature of materials design.

\section*{Conclusion}
To summarize, in this work we have attempted the theoretical evaluation of the pseudo-binary phase diagram of Ti$_2$AlC-Cr$_2$AlC. In particular, first-principles calculations were first conducted to estimate finite-temperature free energies of MAX and competing phases, taking into account both vibrational and electronic contributions. Bayesian CALPHAD was in turn carried out to `refine' the calculated energies in the manner that satisfies the mutual consistency between both quantum-mechanical-based calculations and experimental phase equilibria. It also enables the thermodynamic assessments of phases, that were not assessable by means of first-principles calculations, and more importantly of the uncertainty in model parameters. The calculated energies and their uncertainty were then subjected to a linear minimization to derive the most competitive phases at different temperature-composition conditions. A deterministic phase diagram was in turn constructed based on the results of phase competition. Both agreement and disagreement with experimental phase equilibria have been observed. Except for the cases without crystallographic information, it was found that the disagreement was likely due to the neglect of uncertainty in models and experiments when doing the fully deterministic phase stability evaluation. Uncertainty in model parameters and mass ratio used during synthesis experiments along the pseudobinary Ti$_2$AlC-Cr$_2$AlC system were then accounted for and resulting phase stability showed reasonable agreement with experiments. It is hoped that the current work could point towards a promising way of investigating phase stability in other M$_{n+1}$AX$_{n}$ systems providing the knowledge necessary to elucidate possible synthesis routes for MAX phase systems with unprecedented properties.



\section*{Acknowledgements}
We acknowledge support from NSF through Grants No. DMR-1410983 and CMMI-0953984. AT acknowledges partial support from Grant No. NSF CMMI-1534534. RA acknowledges the support of Texas A\&M's Vice President for Research and Texas Engineering Experiment Station through the Internal Seed Grant on Materials Genomics. First-principles calculations were carried out at the Texas A\& M Super-computing Facility at Texas A\&M University as well as the Texas Advanced Computing Center (TACC) at the University of Texas at Austin. Preparation of the input files and analysis of the data have been performed using AFLOW \cite{2012Curtarolo}. The Texas A\&M Materials Modeling Automation Library (tammal) developed by R.A., A.T. and collaborators (soon to be released to the general scientific community) was used to carry out the HT calculations. The ATAT package \cite{2002VandeWalle} was used to evaluate phonon density required for the calculations of finite-temperature free energies.

\section*{Author contributions statement}

The phase-diagram idea was formulated by T.D., A.T, M.R., and R.A. W.S. contributed to discussions. T.D. and A.T. conducted calculations and analyzed the results. All authors contributed to writing the paper.

\section*{Additional information}

\textbf{Competing financial interests:} The authors declare no competing financial interests.

\end{document}